\documentclass[a4paper,12pt,english]{article}
\usepackage{amsmath}
\title{\textbf{SU(2) Skyrme Model for Hadron}\footnote{\textit{Presented at the Workshop on Theoretical
Physics 2004 (WTP2K4), University of Indonesia, Depok, Indonesia, 19th May 2004.}}}
\author{\textbf{Miftachul Hadi}$^1$,~~~\textbf{Hans Jacobus Wospakrik}$^2$\\\\
$^1$Applied Mathematics for Biophysics Group\\
Physics Research Centre, Indonesian Institute of Sciences (LIPI)\\
Puspiptek, Serpong 15314, Tangerang, Banten, Indonesia\\
E-mail: miftachul.hadi@lipi.go.id\\\\
$^2$Department of Physics, Institut Teknologi Bandung\\
Jl. Ganesha 10, Bandung 40132, Indonesia}
\date{}
\begin{document}
\maketitle
\abstract{The SU(2) Skyrme model is reviewed. The model, which considers hadron as soliton (Skyrmion), is
used for investigating the nucleon mass and delta mass.}

\section{INTRODUCTION} 
A solitary wave was discussed for the first time in 1845 by John Scott Russell in the "Report of the British
Association for the Advancement of Science". Russell observed a solitary wave which was traveling along
Edinburgh-Glasgow water channel, evoluted translationally without any changing in shape and speed for
long enough distance. 

This phenomenon was passed without explanation in Russell life time, until D.J. Korteweg and de Vries gave a complete account of solutions to the nonlinear hydrodynamical equation in 1895. It needs a long time, until Zabusky and Kruskal discovered (which was named after that time as) \textit{soliton} from Korteweg-de Vries equation in 1965 \cite{sascha},\cite{soewono}.

\subsection{Soliton}
\textit{A solitary wave (soliton)} is defined as classical solution of a nonlinear wave equation which has \textit{finite energy}, \textit{nonsingular (noninfinitive) energy density}, \textit{a spatially confined (localized)}, \textit{nondispersive} and \textit{stable properties}. Every soliton is characterized by its \textit{topological invariance} which shows its \textit{stability} \cite{sascha},\cite{hans1},\cite{bala}. 

The importance of soliton was recognized in quite different areas of physics, e.g., information technology obtains the benefit from the use of nondispersive pulses. In particle physics, a localized and stable wave might be a good model for elementary particle. In recent time (2003), the discovery of \textit{pentaquark} enhance the belief that "there is something which can be explored deeper" in soliton as a interesting model of elementary particle \cite{sascha},\cite{soewono},\cite{hans2}.

\subsection{Hadron as a Soliton}
In order to understand the charge radius of nucleon which have size roughly 1 Fermi, T.H.R. Skyrme in
1962 proposed idea that \textit{strongly interacting particles (hadrons) were locally concentrated static solution of
extended nonlinear sigma (chiral) model} \cite{hans3}. 

In (3+1)-dimensions of space-time, it will be observed that Skyrme model tries to explain hadron as soliton from nonlinear sigma (chiral) model field theory in internal symmetrical group of SU(2). 

Skyrme's idea is unifying boson and fermion in a common framework which provides a fundamental field model consist of just pion. The nucleon is obtained, as a certain classical configuration of the pion field. \textit{Skyrmion} is a topological soliton object which is solution of the classical field equation with has localized energy density.

\section{SU(2) SKYRME MODEL}
SU(2) Skyrme model is very simple model, because it only consists of two flavours. This model is described by $U=U(x^\mu)$ function which has SU(2) group valued of (3+1)D space-time coordinates, $x^\mu=(x^0,x^a)$ where $x^0=ct$ is time coordinate, $c$ is light velocity, and $x^a=(x^1=x,~x^2=y,~x^3=z)$ is space coordinate.

The dynamics is determined by \textit{action}
\begin{equation}\label{1}
S=\int d^4x~\mathcal{L},
\end{equation}
where
\begin{equation}\label{2}
\mathcal{L}=\text{Tr}\left[-\frac{F^2}{16}L_\mu L^\mu+\frac{1}{32a^2}\left[L_\mu,L_\nu\right]\left[L^\mu,L^\nu\right]+\frac{F^2}M_\pi^2(U^{-1}+U-2I)\right]
\end{equation}
is the corresponding flat space-time \textit{Lagrangian density} and
\begin{equation}\label{3}
L_\mu=U^{-1}\partial_\mu U
\end{equation}
are \textit{left chiral currents}, $\mathit{F}\cong{123}$ MeV is pion decay constant and $a$ is dimensionless constant. The convention for the space-time metric is $ds^2=(dx^0)^2-(dx^a)^2$.

The first term in equation (\ref{2}) is the SU(2) chiral model Lagrangian density, the second term is \textit{Skyrme term} for stabilizing solitonic solution. The last term descibes the mass term where $\mathit{M}_\pi$ is pion (meson) mass.

Euler-Lagrange equation of SU(2) Skyrme model is derived from \textit{least action principle} 
\begin{equation}\label{4}
\delta S = 0.
\end{equation}
If we take \textit{variation of action} to equation (\ref{2}), we obtain
\begin{equation}\label{5}
\delta S=\int d^4x~\delta\mathcal{L}
\end{equation}
where
\begin{equation}\label{6}
\delta\mathcal{L} = \text{Tr}\left[-\frac{F^2}{8}(\delta L_\mu )L^\mu+\frac{1}{16a^2}\left(\delta \left[L_\mu,L_\nu \right]\right)\left[L^\mu,L^\nu\right]+\frac{F^2}{16}M_\pi^2\left(\delta U^{-1}+\delta U\right)\right],
\end{equation}
\begin{equation}\label{7}
\delta U^{-1}=-U^{-1}(\delta U)U^{-1},
\end{equation}
\begin{equation}\label{8}
\delta L_\mu=-U^{-1}(\delta U)L_\mu+L_\mu U^{-1}(\delta U)+\partial_\mu(U^{-1}\delta U),
\end{equation}
\begin{equation}\label{9}
\text{Tr}~((\delta L_\mu)L^\mu)=-\text{Tr}\left[{(\partial_\mu L^\mu)U^{-1}\delta U}\right]+~\text{total divergence term.}
\end{equation}
Because of $L_{\mu}$ satisfy Maurer-Cartan equations, we can obtain that
\begin{equation}\label{10}
\partial_\mu L_\nu-\partial_\nu L_\mu=-\left[{L_\mu,L_\nu}\right],
\end{equation}
\begin{equation}\label{11}
\delta\left[{L_\mu,L_\nu}\right]=-\partial_\mu\delta L_\nu+~\partial_\nu\delta L_\mu.
\end{equation}
We can derive from (\ref{10}) and (\ref{11}) to obtain
\begin{eqnarray}\label{12}
\text{Tr}\left(\left(\delta \left[{L_\mu,L_\nu}\right]\right)\left[{L^\mu,L^\nu}\right]\right)&=&2~\text{Tr}(\left( {\partial_\mu \left[{L_\nu,\left[{L^\nu,L^\mu}\right]}\right]}\right)U^{-1}\delta U)\nonumber\\
&&+~\text{total divergence term.}
\end{eqnarray}
Substituting (\ref{9}),(\ref{12}) into (\ref{5}),(\ref{6}) and then transform
the total divergence term into surface integral by using the requirement that variation $\delta U=0$ on the volume
boundary of integration
\begin{equation}\label{13}
\delta S=\int{d^4}x~\text{Tr}\bigg[\bigg(\frac{{F^2}}{8}\partial_\mu L^\mu+\frac{1}{{8a^2}}\partial_\mu\left[{L_\nu,\left[{L^\nu
,L^\mu}\right]}\right]+\frac{{F^2}}{{16}}M_\pi^2(U-U^{-1})\bigg)U^{-1}\delta U\bigg].
\end{equation}
From (\ref{4}),(\ref{13}) and because of $\delta U$ has arbitrary value, then Euler-Lagrange equations for Skyrme model Skyrme in matrix form are
\begin{equation}\label{14}
\partial_\mu\left({L^\mu-\frac{1}{{a^2F^2}}\left[{L_\nu,\left[{L^\mu,L^\nu}\right]}\right]}\right)+\frac{1}{2}M_\pi
^2(U-U^{-1})=0.
\end{equation}

\section{STATIC ENERGY}
It will be observed only for static case, where the left chiral currents $L_{\mu=0}=0$.

\subsection{Static Energy and Static Field Equations}
The energy of SU(2) Skyrme model for static case is derived from the corresponding energy-momentum
tensor $T^{\mu\nu}$, which the corresponding energy density is given by $T^{00}$. Explicitly,
\begin{eqnarray}\label{15}
T^{\rho\sigma}
&=&-g^{\rho\sigma}-\frac{{F^2 }}{8}~g^{\mu\rho}g^{\sigma\nu}~\text{Tr}~(L_\mu L_\nu)+\frac{1}{{16a^2}}~(g^{\mu\rho}g^{\sigma\alpha }g^{\nu\beta}\nonumber\\
&&+~g^{\mu\alpha}g^{\nu\rho}g^{\sigma \beta})~\text{Tr}~\left({\left[{L_\mu,L_\nu}\right]\left[{L_\alpha,L_\beta}\right]}\right)
\end{eqnarray}
let $g_{\mu\nu}$ be the metric tensor of curved space-time.

The energy of SU(2) Skyrme model in general (nonstatic case) is
\begin{equation}\label{16}
E =\int{d^3}x~T^{00}
\end{equation}
where $T^{00}$ is \textit{energy-momentum tensor}. Explicitly 
\begin{eqnarray}\label{17}
E&=&\int d^3x~\text{Tr}\left[-\frac{F^2}{16}L_aL_a-\frac{1}{32a^2}[L_a,L_c][L_a,L_c]\right.\nonumber\\&&
\left.
+\frac{F^2}{16}L_o L_o+\frac{1}{16a^2}[L_o,L_a][L_o,L_a]-\frac{F^2}{16}M_\pi^2(U^{-1}+U-2I)-\frac{F^2}{8}L_o L_o\right.\nonumber\\&&
\left.-\frac{1}{8a^2}[L_a,L_o][L_a,L_o]\right]\nonumber\\
&=&E_{\text{static}}+E_{\text{rotation}}
\end{eqnarray}
where,
\begin{equation}\label{18}
E_{\text{static}}=-\int{d^3x~\text{Tr}\left[{\frac{{F^2}}{{16}}L_a^2+\frac{1}{{32a^2}}\left[{L_a ,L_c}\right]^2+\frac{{F^2
}}{{16}}M_\pi^2 (U^{-1}+U-2I)} \right]}
\end{equation}
and
\begin{equation}\label{19}
E_{\text{rotation}}=-\int{d^3x~\text{Tr}\left[{\frac{{F^2}}{{16}}L_0^2+\frac{1}{{16a^2}}\left[{L_0,L_a}\right]^2}\right]}.
\end{equation}

Solitonic properties of SU(2) Skyrme model for static energy in equation (\ref{18}) is studied by scaling the spatial coordinates 
\begin{equation}\label{24}
x\rightarrow2\widetilde{x}/aF 
\end{equation}
and express energy in $F/4a$, i.e. by taking 
\begin{equation}\label{25}
(F/4a)=(1/12\pi^2). 
\end{equation}
In this unit, equation (\ref{18}) becomes
\begin{equation}\label{26}
E_{\text{static}}=\frac{1}{{12\pi^2}}\int{d^3}x\left(-\frac{1}{2}\right)\text{Tr}\left[{L_a^2+\frac{1}{8}\left({\left[{L_a,L_c}\right]^2+m_\pi^2(U^{-1}+U-2I)}\right)}\right]
\end{equation}
where
\begin{equation}\label{27}
m_\pi=2M_\pi/aF.
\end{equation}
Euler-Lagrange equation (\ref{14}) in static case becomes 
\begin{equation}\label{28}
\partial_a\left({L_a-\frac{1}{4}\left[{L_c,\left[{L_a,L_c}\right]}\right]}\right)-\frac{{m_\pi^2}}{2}(U-U^{-1})=0.
\end{equation}

\subsection{Scale Stability}
Let us look at scale transformation below
\begin{equation}\label{29}
x\rightarrow \lambda x.
\end{equation}
We find that $\mathit L_a$ currents, by scale transformation (\ref{29}), transform into
\begin{equation}\label{30}
L_a(x)\to U^{-1}(\lambda x)\frac{\partial U(\lambda x)}{\partial x^a} =
\lambda L_a (\lambda x).
\end{equation}
The effect of scale transformation to static energy (\ref{26}), by ignoring pion mass term (because pion mass is small), is
\begin{equation}\label{31}
E\left[\lambda\right]_{\text{static}}=\frac{1}{\lambda}E_\sigma+\lambda E_{\text{Sky}}
\end{equation}
where $E_\sigma$ is \textit{static chiral energy term} and $E_{\text{Sky}}$ is \textit{Skyrme energy term}.

From equation (\ref{31}), we obtain
\begin{equation}\label{32}
\left.\frac{dE\left[\lambda\right]}{d\lambda}\right|_{\lambda=1}
=\left.\left(-\frac{1}{\lambda^2}E_\sigma+E_{\text{Sky}}\right)\right|_{\lambda=1}
=-E_\sigma+E_{\text{Sky}},
\end{equation}
and
\begin{equation}\label{33}
\left.\frac{d^2 E\left[\lambda\right]}{d\lambda^2}\right|_{\lambda=1}
=\frac{2}{\lambda^3}\left.{E_\sigma}\right|_{\lambda=1}=2E_\sigma.
\end{equation}
The requirement for \textit{extremum stable condition} is
\begin{equation}\label{34}
\frac{{dE\left[ \lambda  \right]}}{{d\lambda }} = 0.
\end{equation}
We apply extremum stable condition, (\ref{34}), to equation (\ref{32}), we obtain
\begin{equation}\label{35}
E_\sigma=E_{Sky},
\end{equation}
which it shows 
\begin{equation}\label{36}
E_{\sigma}\geq 0.
\end{equation}
So that equation (\ref{36}) fulfills condition
\begin{equation}\label{37}
\frac{{d^2E\left[\lambda\right]}}{{d\lambda^2}}> 0.
\end{equation}
Equation (\ref{37}) is \textit{minimum stable condition} which implies that \textit{static energy (\ref{31}) is stable against scale perturbation}.

\section{TOPOLOGICAL CHARGE}
Static energy of SU(2) Skyrme model can be expressed, by ignoring pion mass, as 
\begin{equation}\label{38}
E_{\text{static}}=\frac{1}{{12\pi^2}}\int{d^3}x\left(-\frac{1}{2}\right)\text{Tr}\left[{\left({L_a\pm \frac{1}{4}\epsilon_{abc}\left[{L_b,L_c}\right]}\right)^2}\right]\mp\frac{1}{{24\pi^2}}\int {d^3}x~\epsilon_{abc}\text{Tr}~[L_aL_bL_c]
\end{equation}
where $\epsilon_{abc}$ is Levi-Civita symbol, $\epsilon_{abc}=\delta^{abc}_{123}$. The first integral is positive definite and at the energy lower bound
\begin{equation}\label{39}
E_{\text{static}}\geq B,
\end{equation}
where
\begin{equation}\label{40}
B=-\frac{1}{{24\pi^2}}\int{d^3 x\epsilon _{abc}}\text{Tr}~(L_aL_bL_c).
\end{equation}
$B$ integral, (\ref{40}), is independent of space-metric tensor, i.e. topological quantity which is known as \textit{topological charge} of SU(2) Skyrme model. From (\ref{39}), it can be read that static energy is measured in
the topological charge unit.

\section{STATIC SKYRME AND ITS SOLUTION}
In spherical coordinate, $(r,\theta,\phi)$, static Skyrme equation (\ref{28}) has the following form 
\begin{equation}\label{41}
\begin{split}
0
&=\partial_r\left(L_r-\frac{1}{4}\left\{\frac{1}{r^2}[L_\theta,[L_r,
L_\theta]]+\frac{1}{r^2\sin^2\theta}[L_\phi,[L_r,L_\phi]]\right\}\right)\\
&+\partial_\theta
\left(\frac{1}{r^2}L_\theta-\frac{1}{4}\left\{\frac{1}{r^2}[L_r,[L_\theta,L_r]]+\frac{1}{r^4\sin^2\theta}[L_\phi,[L_\theta,L_\phi]]\right\}\right)\\
&+\partial_\phi\left(\frac{1}{r^2\sin^2\theta}L_\phi-\frac{1}{4}\left\{\frac{1}{r^2\sin^2\theta}[L_r,[L_\phi,L_r]]+\frac{1}{r^4\sin^2\theta}[L_\theta,[L_\phi,L_\theta]]\right\}\right).
\end{split}
\end{equation}
Its solution takes form
\begin{equation}\label{42}
U(r)=\exp(iF_a(r,\theta,\phi)\sigma_a),
\end{equation}
where $U$ is $2\times 2$ unitary matrix, $\sigma_a$ is Pauli matrix, $a=1,2,3$. 
\begin{equation}\label{43}
F_a(r,\theta,\phi)=g(r)n_a(\theta,\phi)
\end{equation}
is \textit{Skyrme ansatz}, $g(r)$ is \textit{profile function} which has spherically symmetric property, and
\begin{equation}\label{44}
n_a=n_a(\theta,\phi),~~~a=1,2,3 
\end{equation}
is component of unit vector, $\widehat{\boldsymbol{n}}$, in the internal space of SU(2), where
\begin{equation}\label{45}
|\hat{n}|^2=n_1^2+n_2^2+n_3^2=1.
\end{equation}
Explicitly, three components of vector $\hat{n}$ are
\begin{eqnarray}\label{46}
n_1&=&\sin\theta~\cos\phi\\
n_2&=&\sin\theta~\sin\phi\\
n_3&=&\cos\theta.
\end{eqnarray}

\section{SPHERIC STATIC ENERGY}
As a result of scaling spatial coordinates, by ignoring pion mass, static energy of SU(2) Skyrme model (\ref{26}) can be stated as 
\begin{eqnarray}\label{47}
E_{\text{static}}=\frac{1}{12\pi^2}(4\pi)\int r^2 dr\left[\left(\frac{dg}{dr}\right)^2+\frac{2}{r^2
}\sin^2g\left(1+\left(\frac{dg}{dr}\right)^2\right)+\frac{1}{r^4}\sin^4g\right].
\end{eqnarray}
where mass term is ignored (very small value). Euler-Lagrange equations for profile function can be derived from equation (\ref{47}), by using least action principle
\begin{equation}\label{48}
\delta_g E_{static} = 0,
\end{equation}
we obtain
\begin{equation}\label{49}
\frac{{d^2g}}{{dr^2}}\left[{1+\frac{2}{{r^2}}\sin^2g}\right]
+\left({\frac{{dg}}{{dr}}}\right)^2\left[{\frac{1}{{r^2}}\sin 2g}
\right]+ \left({\frac{{dg}}{{dr}}} \right)\left({\frac{2}{r}}\right)-\frac{1}{{r^2}}
\sin 2g-\frac{1}{{r^4}}\sin^2 g\sin 2g=0.
\end{equation}
Equation (\ref{49}) is second order of nonlinear differential equation. 

The solution of equation (\ref{49}) will be worked numerically with finite difference method which gives the value of $g(r)$. In order that static energy (\ref{47}) has finite value at $r=0$ and $r=\infty$, then profile function $g(r)$ must fulfills \textit{boundary conditions}
\begin{equation}\label{50}
g(0)=\pi,~~~g(\infty)=0.
\end{equation}
If the value of $g(r)$ is obtained then we can be use it to calculate: static energy, static mass of nucleon and delta, Skyrmion moment of inertia.

\section{SKYRMION QUANTIZATION}
In the static case, using Skyrmion ansatz (\ref{43}), (\ref{42}) becomes:
\begin{equation}\label{51}
U(\vec{r})=\exp(ig\sigma_a.\vec{r})
\end{equation}
where
\begin{eqnarray}\label{52}
\hat{n}&=&\hat{r}\\
\hat{r}&=&\vec{r}/|\vec{r}|\\
\hat{r}&=&(x_a/r)\hat{i}
\end{eqnarray}
$i$ is base vector.
Observation of (\ref{51}) to time dependence
\begin{equation}\label{53}
U(\boldsymbol{r})\to U(\boldsymbol{r},t)=A(t)U(\boldsymbol{r})A(t)^\dag,
\end{equation}
\begin{equation}\label{53.1}
A(t)\in SU(2)_{\text{internal}},
\end{equation}
\begin{equation}\label{54}
AA^\dag=A^\dag A=I
\end{equation}
where $A$ is time dependent unitary matrix. $A(t)$ matrix is related to $R(t)$ rotation matrix of space coordinate:
\begin{equation}\label{55}
r\rightarrow r'=R(t)\vec{r}
\end{equation}
which shows that the transformation effect of SU(2) internal is equal to the transformation effect of \textit{space rotation}. $A$ matrix is called \textit{the collective coordinate} of Skyrmion.

\section{QUANTIZED ROTATIONAL ENERGY}
Skyrmion quantization is worked by involving time dependent of $U(\boldsymbol{r})$ as 
\begin{equation}\label{56}
U(\boldsymbol{r})\to U(\boldsymbol{r},t)=A(t)U(\boldsymbol{r})A(t)^\dag.
\end{equation}
From equation (\ref{19}) and by using equation (\ref{56}), rotational energy can be stated as 
\begin{eqnarray}\label{57}
E_{\text{rotation}}
&=&-\left({\frac{{\pi F^2}}{6}\int\limits_0^\infty{dr~r^2}\sin^2g+\frac{{2\pi}}{{3a^2}}\int\limits_0^\infty{dr~r^2\sin^2g\left[{\left({\frac{{dg}}{{dr}}}\right)^2+\frac{1}{{r^2}}\sin^2g}\right]}}\right)\nonumber\\
&&\text{Tr}\left({R^{-1}\frac{{\partial R}}{{\partial t}}}\right)^2\nonumber\\
&=&\frac{1}{2}~I~\text{Tr}~\boldsymbol{\Omega}^2
\end{eqnarray}
where
\begin{equation}\label{58}
\text{Tr}~\boldsymbol{\Omega}^2=-\text{Tr}\left(R^{-1}\frac{\partial R}{\partial t}\right)^2
\end{equation}
with $\boldsymbol{\Omega}$ is angular velocity matrix of Skyrmion,
and
\begin{equation}\label{59}
I=2\left(\frac{\pi F^2}{6}\int^\infty_0dr~r^2\sin^2g+\frac{2\pi}{3a^2}\int^\infty_0 dr~r^2\sin^2g\left[\left(\frac{dg}{dr}\right)^2+\frac{1}{r^2}\sin^2g\right]\right)
\end{equation}
is Skyrmion moment of inertia. Following the classical mechanics, angular momentum of Skyrmion can be stated as
\begin{equation}\label{60}
\vec{J}=\frac{\partial E_{\text{rotation}}}{\partial\vec{\omega}}=I\vec{\omega}
\end{equation}
So, rotational energy of Skyrmion is
\begin{equation}\label{61}
E_{\text{rotation}}=\frac{\vec{J}^2}{2I}.
\end{equation}
Following the quantum mechanics, angular momentum is given by
\begin{equation}\label{62}
\vec{J}^2=j(j+1)\hbar^2 
\end{equation}
where $j= 0,\frac{1}{2},1,\frac{3}{2},2,...$ and $\hbar = h/2\pi$ is Planck constant. It is used natural unit $\hbar = 1$. Then (\ref{61}) becomes:
\begin{equation}\label{63}
E_{\text{rotation}}=\frac{{j(j+1)}}{2I}.
\end{equation}

\section{HADRON MASS}
Nucleon mass (spin $\frac{1}{2}$) is given by the relations
\begin{equation}\label{64}
m_N=M+\frac{j(j+1)}{2I}.
\end{equation}
For $j=\frac{1}{2}$ then (\ref{64}) becomes 
\begin{equation}\label{65}
m_N=M+\frac{\frac{1}{2}(\frac{1}{2}+1)}{2I}.
\end{equation}
where
\begin{equation}\label{66}
M=\frac{E_{\text{static}}}{c^2}.
\end{equation}
$\Delta$ mass (spin $\frac{3}{2}$) is given by
\begin{equation}\label{67}
m_\Delta=M+\frac{j(j+1)}{2I}.
\end{equation}
For $j = \frac{3}{2}$ then (\ref{67}) becomes:
\begin{equation}\label{68}
m_\Delta=M+\frac{\frac{3}{2}(\frac{3}{2}+1)}{2I}.
\end{equation}
Wess-Zumino quantization condition \cite{bala}, requires:
\[j = \frac{1}{2},\frac{3}{2},\frac{5}{2},\ldots\]
It means that Skyrmion is fermion.

\section{SUMMARY}
SU(2) Skyrme model treats hadron as soliton (Skyrmion). The dynamics is shown by Euler-Lagrange equation or Skyrme equation in (\ref{14}), which (\ref{28}) is for static case. The energy in static case is derived from the corresponding energy-momentum tensor. 

Solitonic property of static energy is studied by scaling the spatial coordinates and then examine the scale stability by scale transformation. Stable conditions, i.e. extremum stable condition, (\ref{34}), and minimum stable condition, (\ref{37}), imply that the static energy (\ref{31}) is stable against scale perturbation. 

Numerical solution of (\ref{49}) by using Skyrme ansatz gives the value of profile function, $g(r)$. If the value of profile function is known then it can be used to calculate static energy (\ref{47}), static mass (\ref{66}) and Skyrmion moment of inertia (\ref{59}).

Skyrmion quantization is observed by which the transformation effect of SU(2)$_{\text{internal}}$ is equal to the
transformation effect of space rotation. The nucleon mass (\ref{64}) and $\Delta$ mass (\ref{67}) are the relations of rotational energy (\ref{63}) and static mass (\ref{66}). The work in progress.

\section{ACKNOWLEDGMENTS}
MH thanks to Hans J. Wospakrik, Ph.D for extraordinary supervision and extraordinary patience, Dr. L.T. Handoko for introducing us in elementary particles, early supervision and all kindness, Dr. Terry Mart for very nice lectures on quantum mechanics and nuclear theory, Dr. Freddy P. Zen for introducing string theory and Dr. Agus Purwanto for motivation and inspirations. 

Thank to the Abdus Salam ICTP for Visiting Fellowships of Summer School on Particle Physics 2001, Indonesia Institute of Sciences for Graduate Scholarships, Physics Research Centre LIPI for research facilities, all kindly colleagues for their best supports.

In depth, MH thank to beloved Mother for very long contributions, praying and motivation. Special thank to beloved Ika Nurlaila for huge encouragement and Aliya Syauqina Hadi for much love.

\end{document}